\documentclass[sigconf]{acmart}

\AtBeginDocument{%
  }

\usepackage{booktabs} 
\usepackage{amsmath}
\usepackage{graphicx}
\usepackage{multirow}
\usepackage[caption=false, font=footnotesize]{subfig}
\usepackage{xspace}
\setcopyright{acmcopyright}
\copyrightyear{2018}
\acmYear{2018}
\acmDOI{XXXXXXX.XXXXXXX}

\acmConference[XYZ]{Make sure to enter the correct
  conference title from your rights confirmation emai}{November 2023}{USA}
\acmPrice{15.00}
\acmISBN{978-1-4503-XXXX-X/18/06}

\DeclareRobustCommand{\company}{Company-X\xspace}
\DeclareRobustCommand{\system}{{\sc Kyurem}\xspace}
\DeclareRobustCommand{\basesystem}{{\sc Magneton}\xspace}

\newcommand{\stitle}[1]{\vspace{0.2em}\noindent\textbf{#1}}
\newcommand{\hide}[1]{}
\newcommand{\eg}{{\itshape e.g.}, }
\newcommand{\ie}{{\itshape i.e.}, }

\setcopyright{none}

\begin{document}

\title{Knowledge Acquisition and Integration with Expert-in-the-loop}

\author{Sajjadur Rahman}
\email{sajjadur@megagon.ai}
\affiliation{%
  \institution{Megagon Labs}
  \city{Mountain View}
  \state{California}
  \country{USA}
}
\author{Frederick Choi}
\authornote{Work done during internship at Megagon Labs.}
\email{fc20@illinois.edu}
\affiliation{%
  \institution{University of Illinois at Urbana-Champaign}
  \city{Urbana}
  \state{Illinois}
  \country{USA}
}
\author{Hannah Kim}
\email{hannah@megagon.ai}
\affiliation{%
  \institution{Megagon Labs}
  \city{Mountain View}
  \state{California}
  \country{USA}
}
\author{Dan Zhang}
\email{dan_z@megagon.ai}
\affiliation{%
  \institution{Megagon Labs}
  \city{Mountain View}
  \state{California}
  \country{USA}
}
\author{Estevam Hruschka}
\email{estevam@megagon.ai}
\affiliation{%
  \institution{Megagon Labs}
  \city{Mountain View}
  \state{California}
  \country{USA}
}

\begin{abstract}
Constructing and serving knowledge graphs (KGs)
is an iterative
and human-centered process involving
on-demand programming and analysis.
In this paper, we present \system, a programmable and interactive widget 
library that facilitates human-in-the-loop knowledge acquisition and integration to enable
continuous curation a knowledge graph (KG). 
\system provides a seamless environment within computational notebooks where
data scientists explore a KG to identify opportunities for acquiring new knowledge and 
verify recommendations provided by AI agents for integrating the acquired knowledge in the KG.
We refined \system through participatory
design and conducted case studies
in a real-world setting for evaluation.
The case-studies show that introduction of \system
within an existing HR knowledge graph construction and serving platform
improved the user experience of the experts and helped eradicate inefficiencies
related to knowledge acquisition and integration tasks.

\end{abstract}



\maketitle

\section{Introduction}
With enterprises increasingly adopting Knowledge Graphs (KGs) as the foundation for managing data and building intelligent agents for search, recommendation, and conversation, among others, graph-centered data science is practiced more and more within organizational workflows~\cite{gutierrez2021knowledge}.
These workflows involve iterative and human-centered tasks such as new knowledge acquisition and recommendation, knowledge integration, and curation of the modeled information~\cite{ilyas2022saga, bharadwaj2017creation, knowledgevault, haase2019metaphactory}. 
Therefore, supporting these workflows require experts such as data scientists to blend programming with interactive exploration.
While computational notebooks are suitable for supporting such workflows, the implications of instrumenting programmable and interactive interfaces 
for workflows involving knowledge graphs is underexplored.

\begin{figure*} 
\vspace{-10pt}
    \centering
  \subfloat[\label{fig:overview}]{%
       \includegraphics[width=0.18\linewidth]{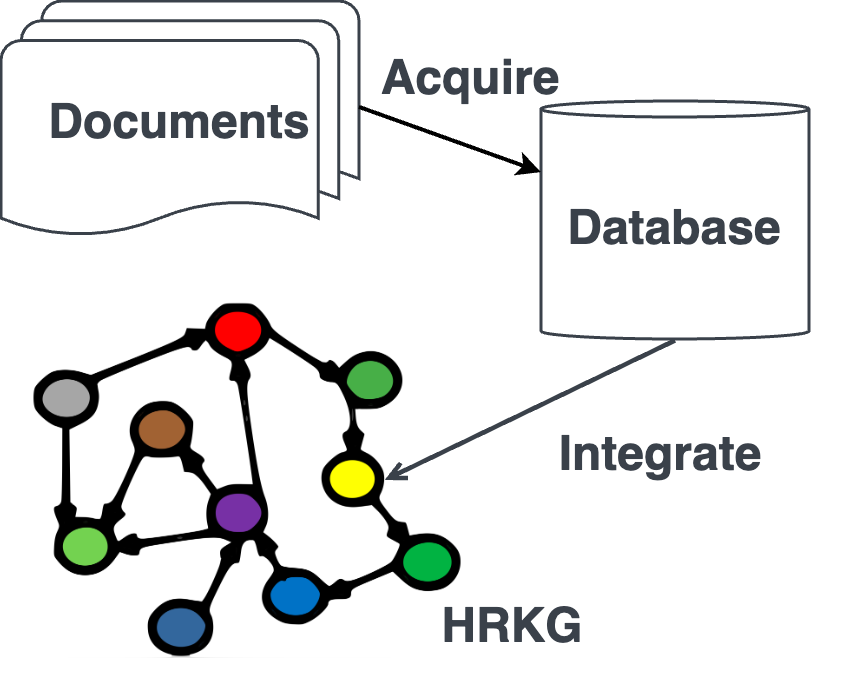}}
    \hfill
  \subfloat[\label{fig:study_flow}]{%
        \includegraphics[width=0.28\linewidth]{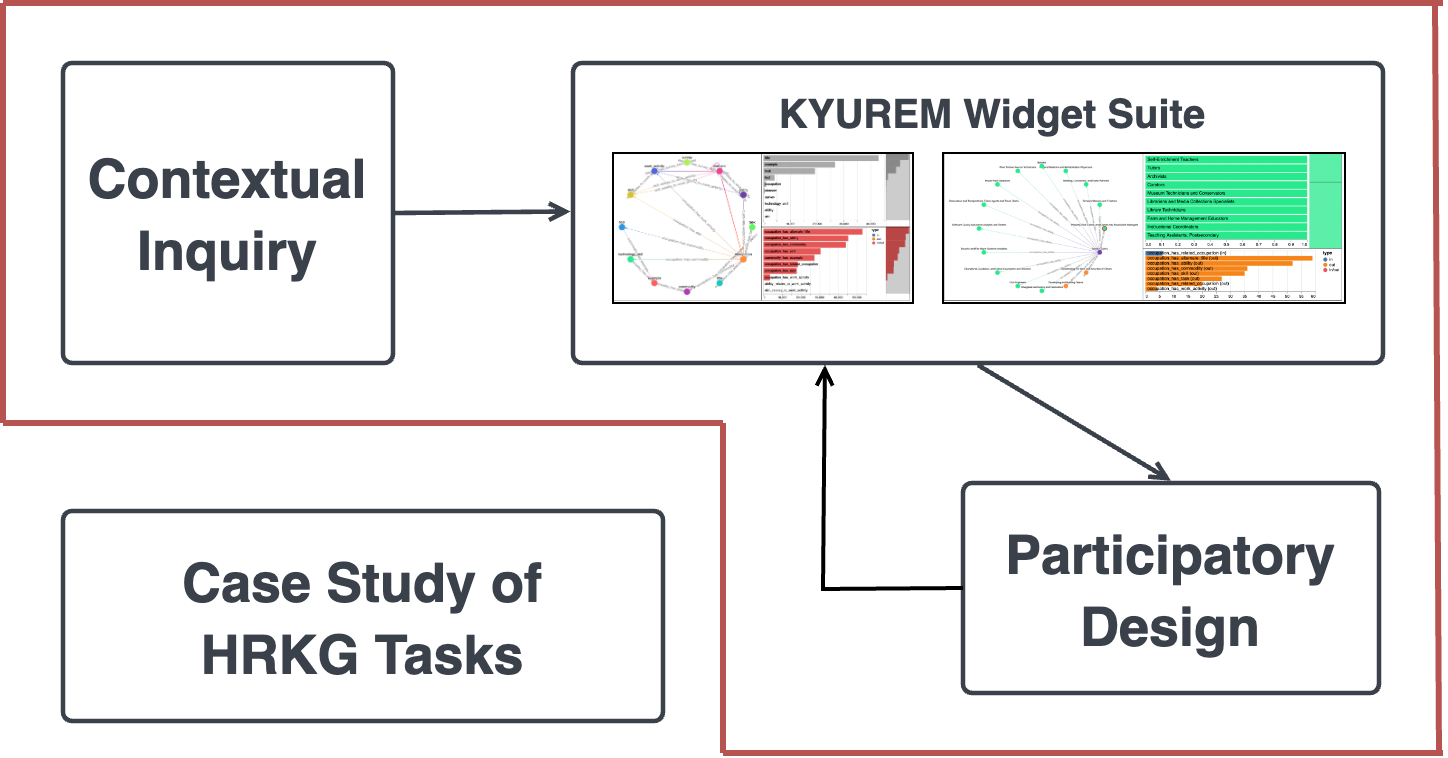}}
    \hfill
  \subfloat[\label{fig:archi}]{%
        \includegraphics[width=0.48\linewidth]{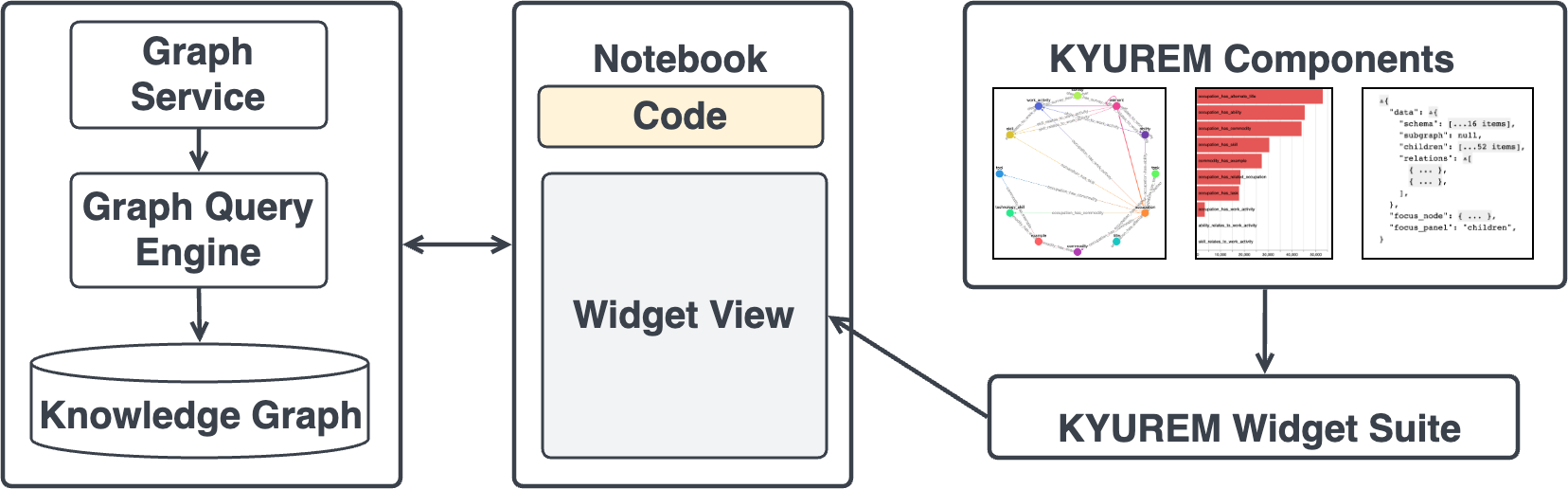}}
  \caption{(a) An overview of the \company HRKG tasks, (b) the study design, and (c) the \system system architecture.}
  \label{fig:overview} 
\vspace{-10pt}
\end{figure*}

In this work, we focus on two specific tasks within the human resources knowledge graph (HRKG hereafter) construction and serving platform at \company --- knowledge acquisition and knowledge integration, an industrial lab focusing on natural language processing, data management, and human-centered AI research. 
\company is a subsidiary of a large holding and conducts research and development for the other subsidiaries with worldwide businesses in staffing, human resources, travel, marketing, and other online consumer services.
Knowledge graph (KG) construction and serving plays a central role at \company. 
Deploying and maintaining a KG that can be shared across 
applications has obvious benefits, as applications such as QA systems, search engines, 
conversational agents~\cite{quamar2020ontology}, and training large language models~\cite{zhao2021knowledge}
require accurate and up-to-date facts about entities and relations. 
Therefore, facilitating a broad range of knowledge that is accurate
and updated continuously is of utmost importance. The HRKG platform
at \company is similar to other such platforms
in industry~\cite{ilyas2022saga, bharadwaj2017creation, knowledgevault, haase2019metaphactory},
where, information about entities is obtained by integrating data from multiple 
structured (\eg relational databases, taxonomies, other KGs) and unstructured sources (\eg documents, web text)~\cite{chen2020kgpt}.

As shown in Figure~\ref{fig:overview}, the KG construction and maintenance involves various tasks from knowledge acquisition to
knowledge integration and canonicalization.
All of these steps involve multiple stakeholders (\eg engineers, researchers, taxonomy architects)
and require significant human intervention. 
Within the KG construction and serving workflow, data practitioners face many challenges. To better understand the challenges and to provide meaningful solutions we conducted several studies within \company (see Figure~\ref{fig:study_flow}.)
We conducted a contextual inquiry-style study~\cite{holtzblatt1997contextual}, where we observed the work practices of three data practitioners working on projects related to knowledge acquisition and integration within the HRKG platform. The study helped understand the pain points of the practitioners and informed design guidelines to assist the experts in performing knowledge acquisition and integration within the HRKG. One such example is the negative user experience with existing tools stemming from frequent context switching to accommodate multiple objectives, \eg programming, and visualization. Using programmable and interactive interfaces integrated with Computational notebooks (\eg Jupyter~\cite{kluyver2016jupyter} and Observable~\cite{observable}) may minimize such context switching.

To address these gaps, we implement \system widget suite --- a library of interactive widgets built using the \basesystem framework~\cite{choi2023towards} --- to specifically support expert-in-the-loop knowledge acquistion and integration tasks. The in-notebook widgets display task-specific visualizations and support user interactions on the interface. The task-specific visualizations and their corresponding interactions are implemented based on design guidelines identified from existing work on graph visualization and exploration~\cite{liu2018graph, 2019_eurovis_mvnv, pienta2015scalable, pretorius2014tasks, kerracher2015task, ahn2013task, lee2006task}. To enable graph-centric data science, we build a python library that translates user interactions on the widgets to graph queries. 
We worked with $12$ data practitioners (\eg researchers, engineers, and data scientists) at \company in weekly participatory design sessions over six weeks to further understand their requirements. Informed by these sessions, we implemented additional \system widgets and components
supporting tasks such as new knowledge acquisition and acquired knowledge integration. Finally, we conducted two case studies to understand the implications of instrumenting \system within the in-house HRKG platform. The experience of the experts-in-the-loop were positively impacted by the introduction of \system. Participants also found new insights that were previously not apparent to them due to the absence of interactive interfaces, such as data quality issues with the knowledge acquisition tasks.

\section{Related Work}
\label{sec:related}
A graph or a network is an abstract data type consisting of a finite set of nodes (entities) and edges (relations)~\cite{liu2018graph}. Depending on the properties of the nodes and edges, graphs can be of various types: homogeneous or heterogeneous and directed or undirected~\cite{liu2018graph, 2019_eurovis_mvnv}. Furthermore, graphs can be static or dynamic. In this paper, we focus on static \emph{Knowledge Graphs} (KGs) --- a directed heterogeneous graph with edges encoding facts among nodes (\eg Person \emph{lives\_in} City)~\cite{kg2019def}. Visualizing KGs requires displaying both graph topology (\ie structure) and attributes associated with the nodes and edges.
We refer the readers to a survey of heterogeneous graph visualization for further details~\cite{2019_eurovis_mvnv}. Besides the node and edge properties, any graph visualization system must consider other aspects such as scale, sparsity, and visualization goal. In \system, we made such rendering decisions by referring to the usage guidelines communicated through the decades-long research conducted by the data visualization and HCI communities on defining graph layouts, encoding, and underlying data operations~\cite{liu2018graph, 2019_eurovis_mvnv, pienta2015scalable}. 
Instead of generating visualizations from scratch, \system leverages existing graph visualization libraries. 
These libraries provide built-in features such as
choice of multiple layout algorithms, scaling to larger datasets, customization and extensibility for add-ons, compatibility with different setups (\eg browsers, touch screens), rendering capabilities in various formats (\eg WebGL, SVG, and CSS-based), and integration with existing graph databases such as Neo4j~\cite{miller2013graph}. We utilize an existing embeddable graph visualization library called Cytoscape.js~\cite{franz2016cytoscape} as \system embeds visualizations in Jupyter Notebook cells. Other popular libraries include
D3.js~\cite{bostock2011d3}, G6~\cite{g6}, Neovis.js~\cite{neovis},  Sigma.js~\cite{sigmajs}, and Vivagraph.js~\cite{vivagraphjs}, among others.

The primary goal of \system is to support various exploration tasks within knowledge graphs as users work on projects related to KG construction and serving. 
To this end, we refer to existing graph task taxonomies to define various interactions representing a user's exploration goal~\cite{pretorius2014tasks, kerracher2015task, ahn2013task, lee2006task}. 
Various bespoke graph exploration tools are available in industry~\cite{graphileon, graphistry, graphpolaris, graphxr, hume, linkurious, neobloom, neodash, yworks}. 
These tools enable graph exploration via direction manipulation or graph querying. These tools do not support blending codes with visualizations, forcing users to use different tools for their analysis workflow.
Graphistry~\cite{graphistry}, igraph~\cite{igraph}, and NetworkX~\cite{networkx} offer python packages and libraries for network analysis in computational notebooks such as Jupyter Notebook. Graphistry additionally supports GPU-accelerated rendering of large graph visualizations. However, the generated visualizations are static objects displayed in notebook cells lacking interactivity and are unsuitable for interactive graph exploration. 
There are academic prototypes for general purpose and targeted graph exploration as outlined in various surveys~\cite{pienta2015scalable,liu2018graph}. 
While these are also bespoke tools lacking integration with interactive programming environments, they inform different layouts for visualizing graphs~\cite{henry2007nodetrix, shneiderman2006network}, interactions~\cite{pienta2017facets, pienta2017vigor, chau2011apolo}, and underlying data operations~\cite{shen2006visual}. 
\system, on the other hand, is embedded within the existing data science ecosystem, \ie in computational notebooks. 

\section{Design Goal Formulation}
\label{sec:design_goal}
We now present a contextual inquiry of data science workflows within the HRKG platform 
that helped understand the challenges of the practitioners with their existing workflows and informed design guidelines for building assistive tools.

\subsection{Contextual Inquiry of Graph Exploration Workflows}
Contextual inquiries are ethnographic field studies where researchers interview a small group of people in their natural environment as they conduct their work activities (\textit{context}) to gain an in-depth understanding of their work practices~\cite{holtzblatt1997contextual}. 
As opposed to formative studies which are reflective and require participants to recall their experiences from memory, contextual inquiries take place in the participants' day-to-day work environment --- \company in this case.
We conducted the study by recruiting three data practitioners --- one researcher, one engineer, and one data scientist --- by reaching out to them in Slack\footnote{\url{https://slack.com/}}. 
These participants contributed to various projects related to knowledge acquisition (discovering new entities and relations), curation (graph understanding and standardization), and integration (graph alignment and merging) within the HRKG platform. Participants in each session brought their laptops, tools, and documents (Google Sheets and Slides).
The sessions lasted for $45$ minutes, where we observed participants performing their project-specific tasks with their existing tools and setup.
During observation, we intervened as needed to ask follow-up questions to gain a deeper understanding of the participants'
decisions and their pain points. 




\subsection{Key Takeaways}
\stitle{Tedious Context Switching.} While the participants were involved in different projects within the platform, such as knowledge acquisition, curation, and integration, their work setup involving graphs was somewhat similar. These individual projects or phases could be further broken down into smaller tasks ---  viewing and assessing data, formulating a hypothesis on how to perform the intended task, and verifying the outcomes of the executed task --- similar to Rahman et al.~\cite{rahman2022ie}. However, completing such multi-purpose tasks required participants to use different tools --- command-line environments, computational notebooks such as Jupyter, and code editors such as Visual Studio Code\footnote{https://code.visualstudio.com/} for programming; Neo4j browser\footnote{\url{https://neo4j.com/developer/neo4j-browser/}} for querying and visually exploring graphs and tools such as Matplotlib~\cite{hunter2007matplotlib} for visualizing data;  editors such as spreadsheets, and notebook cell outputs for viewing and verifying program outputs. Participants complained about tedious context switching among these tools. 

\stitle{Lack of Flexibility in Workflows.} Surprisingly, participants' graph exploration experiences were limited to using the Neo4j browser. While there are bespoke tools~\cite{graphistry, graphileon,linkurious} and even Neo4j desktop extensions~\cite{neobloom, neodash,yworks}, using those meant (a) ``\emph{adding another new tool}'' to the already iterative and messy experience and (b) learning a new platform. Moreover, using desktop-based tools had additional limitations, \eg company policy prohibiting the hosting of proprietary and sensitive data on personal computers. 
When using the Neo4j browser the participants could not leverage rich graph layouts, visualizations, and interactions available in various charting libraries mentioned in Section~\ref{sec:related}. Additionally, the browser lacked support for common visualizations such as bar charts and scatter plots, which are essential for downstream analysis and decision making. For viewing these charts, participants used a combination of notebooks or code editors and charting libraries in Python.

\subsection{Design Goals}
We opted for developing \system, an interactive widget library designed to specifically support the knowledge curation and integration tasks within computation notebooks. Through \system we aim to support the rendering of graph visualizations in notebooks and provide libraries that enable users to execute a suite of graph-specific data operations for visualizing and exploring graphs. \system is built on top of the \basesystem ~\cite{choi2023towards}, an extensible framework which enables data science workflows within interactive programming environments. Our choice was informed by the observations of the contextual inquiry and solidified by the recommendation by existing work to blend the expressivity of codes with the interactivity of visualizations to ensure flexibility in workflows~\cite{futzing2019moseying} and evaluation studies conducted in recent work highlighting the suitability of such environments in eliminating context switching~\cite{wu2020b2, bauerle2022symphony}. 





\section{\system Design}
\label{sec:system}
We now provide an overview of \system and it's architecture and discuss how it evolved via participatory design. 


\subsection{Participatory Design}

Inspired by earlier work~\cite{bauerle2022symphony, rahman2020mixtape}, we conducted several participatory design sessions after developing the initial \system framework. These sessions aimed to understand the specific needs of data science workers and then design and develop a core set of \system components and widgets. We conducted weekly sessions over a six-week period focusing on three different projects related to knowledge acquistion and integration. These projects involved several collaborators in various roles within \company 
from researchers ($N=6$) 
and engineers ($N=4$) 
to product managers ($N=1$) 
and taxonomy architects ($N=4$). 
Note that the taxonomy architect role was shared by a subset of 
researchers and engineers within the platform team.
The sessions lasted between 30 minutes and an hour.
The first session with each team involved introducing the core features of \system and the base components. We then asked participants to describe their tasks step by step.
While the weekly sessions involved the entire platform team, 
we also arranged ad-hoc participatory sessions
with individual project leaders for a deeper dive into 
the project-specific requirements and 
iteratively refined \system features.


\begin{figure*}[!htb] 
  \centering
  \includegraphics[width=\linewidth]{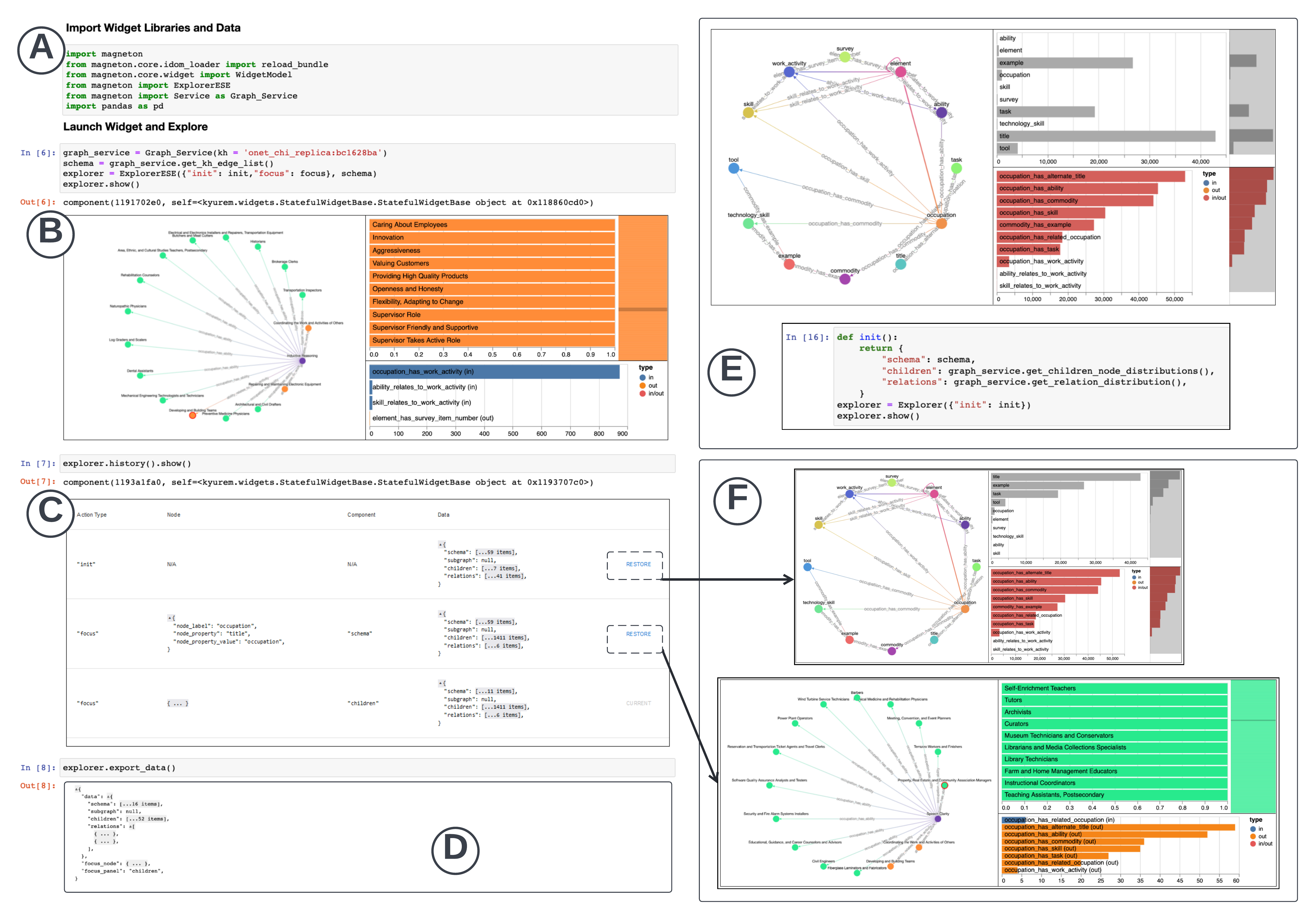}
  \caption{Screenshot of a knowledge acquisition workflow using \system widget library. (a, b) importing libraries to launch a multiple-coordinated view widget, (c) leveraging the history view of \basesystem to context switch between exploration history, (d, e, f) viewing and exploring multi-modal data interactively and programmaticially.}
  \label{fig:alignment-verification} 
\end{figure*}

We collected several systems- and feature-level enhancement requests for \system from these participatory design sessions. These requests fall into three themes: data modality (\ie ability to explore multi-modal data such as text, JSON documents, and graphs), perceptual scalability (\ie ease of overviewing large-scale data), and synchrony of navigation (\ie seamless context switching among multiple-coordinated views.) Based on the participant feedback, we implemented eight basic components shown in Figure~\ref{fig:defaults} in Appendix~\ref{app_sec:components}. These components were designed to support various enhancements and augmentation requests 
mentioned above: multi-modal data exploration (components b,f,g, and h in Figure~\ref{fig:defaults}),
(ii) scalable overview via faceted graph (component a in Figure~\ref{fig:defaults}), and (iii) multiple-coordinated views achieved by combining components into widgets (Figure~\ref{fig:ese-explorer}.)
We provide a detailed account of how the participatory session with participants helped prioritize the select of these components in Appendix~\ref{app_sec:components}. 

\subsection{\system System Architecture}
Figure~\ref{fig:archi} provides an overview of the system we developed to support graph exploration and analysis using the \system library. At the heart of the system is the \system widget suite, a collection of widgets that support programmatic and interactive graph exploration within computational notebooks. The base widget in the suite is extended from the \basesystem framework~\cite{choi2023towards}. Users' actions in the front-end components trigger various functions, via Graph Service, in the back-end.
Examples of graph-specific operations include generating graph schema, finding node neighborhoods, and computing distributions of nodes and relations, among others. The operations are part of an in-house graph query library, which is published as a Python package. Each operation is mapped to the corresponding graph query, which is executed by the graph query manager and returned to the front-end. Note that the graph data operations can be initiated from the widget, notebook, or the front-end components.

\section{Case Study}
\label{sec:study}
We performed a preliminary evaluation of \system with data practitioners working on two tasks within the HRKG platform: new knowledge acquisition and knowledge integration.  
To understand the usage benefits and limitations of \system, 
we conducted case studies lasting 60 minutes where a member of each
the team employed \system to accomplish their respective tasks in a Jupyter notebook.
For the study, the participants worked with an HR knowledge graph and a related job description corpus. However, the case studies involved \company's proprietary knowledge graph and datasets. Therefore, to demonstrate various features of \system, we use a knowledge graphs and dataset constructed from O*NET~\cite{peterson1999occupational}. However, the information conveyed in the screenshots reflects the original work setting and experience of the participants.
We asked the participants about their experience using \system widgets and discussed the limitations of \system within the context of their tasks and the HRKG platform. Note that the interviews were recorded (with participant permission), and transcripts were generated via an automated transcription service. We further corrected spelling and grammatical mistakes in those documents by referring to the recording. Therefore, the quotes presented in this paper are essentially paraphrases. We now discuss the case-study scenarios in detail.

\subsection{Study 1: New Knowledge Acquisition}
\begin{figure*}[!htb] 
  \centering
  \includegraphics[width=0.8\linewidth]{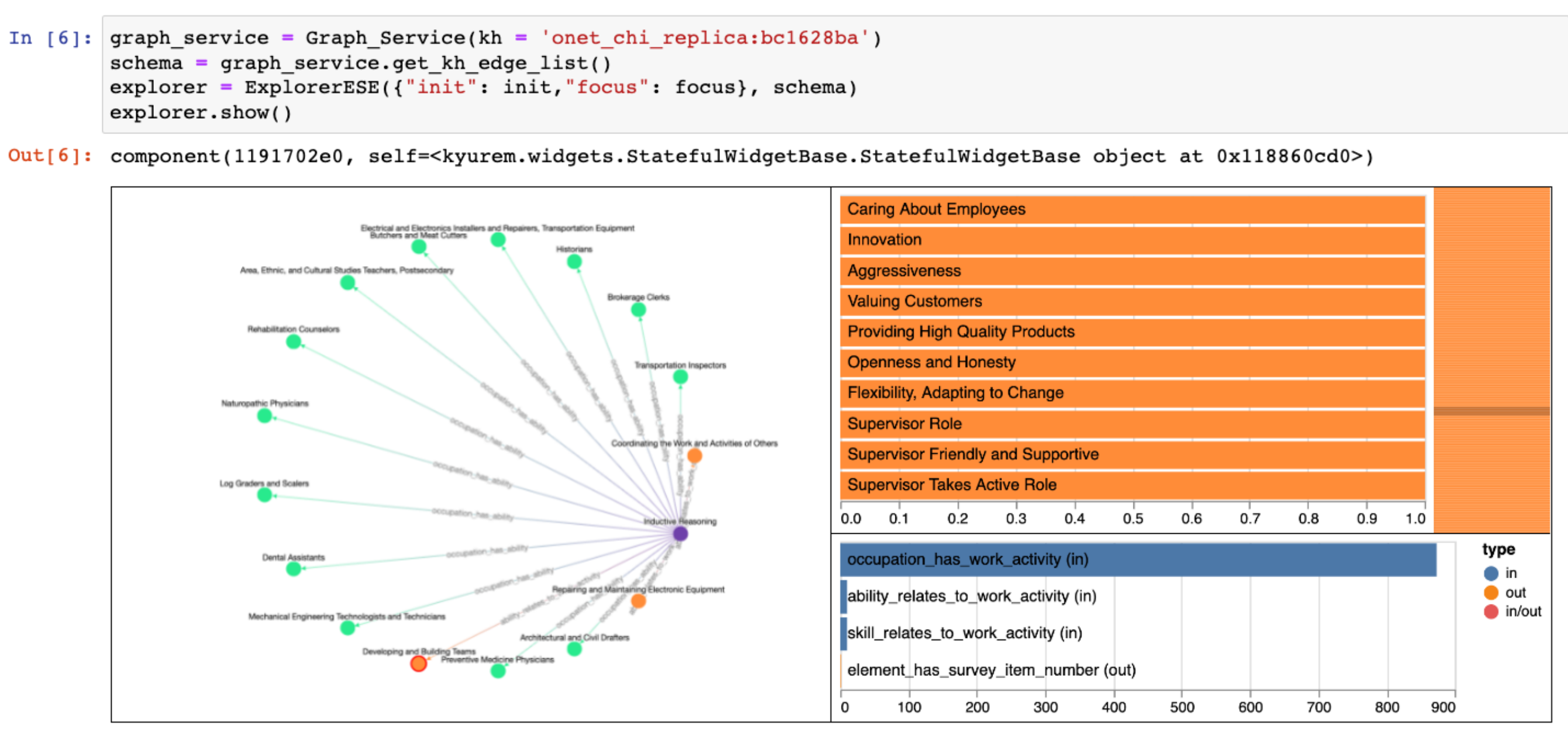}
  \caption{Knowledge acquisition widget: ESE-explorer. Users can launch the interactive widget from a Jupyter notebook, select seed examples of a concept, and export the seeds programmatically.}
  \label{fig:ese-explorer} 
\end{figure*}

The participant for this case study worked in the knowledge acquisition project deployed to add new entities and relations to the graph. The newly discovered entities and relations expand the knowledge graph, further enhancing the domain understanding.  
For this session, the participant only focused on a new entity mining task known as entity set expansion (ESE)~\cite{shao-etal-2022-low}. 
Given a collection of user-provided examples, known as \emph{seeds} (\eg inductive reasoning) of an entity type (\eg ability), an ESE algorithm mines new instances of that entity type from a text corpus (\eg job descriptions). In this study, the participant focused on exploring the graph to identify potential candidates (entity types) for expansion and their corresponding seed, \ie example entities. Furthermore, the participant wanted to explore the mention of an entity in a job description corpus to gauge their suitability as seeds for the ESE task subjectively. Given these requirements, we provided an \emph{ESE-explorer} widget (Figure~\ref{fig:ese-explorer}) by extending the explorer widget used in the first study. We added a static table component (Figure~\ref{fig:defaults}h) to view the text corpus.

\stitle{Elimination of Context Switching.} When exploring the notebook, the participant found the coordinated views between the sub-graph and table components to be the most useful. In their previous workflow, participants explored the graph (to collect seeds) and the corpus (to verify seed suitability) in the Neo4j browser and spreadsheets, respectively. The exercise forced the participant to move back and forth between multiple tools. The other alternative was to write multiple scripts, one for querying graph nodes (using Cypher) and another for searching and returning matching job descriptions (using Python), and then view the results in a spreadsheet. The experience was so cumbersome that the participant would be left ``exhausted'' due to frequent tool switching.

\stitle{Easier Qualitative Validation.} The exploration in the \system widget helped the participants make informed decisions and even discard candidates that they previously selected based on commonsense --- ``looking at the context, I would not have selected this (entity) as a seed.'' Moreover, the distribution component helped steer the participant's exploration and find suitable entity types for expansion --- low-frequency entities in the graph are good candidates for further expansion based on an external source. The participant was able to explore the corpus mention of entities of a given entity type to subjectively measure their expansion potential --- ``judging by their use in the corpus, this (entity type) does not seem like a good enough concept to grow.''

\subsection{Study 2: Knowledge Integration}
In the other case study, the participant was involved in the knowledge integration project, which focused on aligning and merging external 
information in the knowledge graph. Examples include entity alignment~\cite{sun2020benchmarking} between two knowledge graphs and aligning entities from the text corpus into the graph. In this session, the participant focused on verifying potential alignment candidates extracted from a text corpus and assigning merging decisions by exploring candidate entities in the in-house knowledge graph. We provided an \emph{verification} widget (Figure~\ref{fig:alignment-verification} in Appendix~\ref{app_sec:components}) for the verification task. The widget contained an interactive table component (Figure~\ref{fig:defaults}h) with a selection menu of alignment decisions (\eg insert, ignore, defer) in one of the columns. The other two columns contained alignment candidates from two different sources: a job description corpus and a knowledge graph. Clicking a candidate displayed their context in the corresponding source which was displayed using the static table component. At the same time, a sub-graph component would display the corresponding graph entity neighborhood. 



\stitle{Seamless Workflow Execution.} The participant's previous experience was to run alignment scripts in a command-line interface and then view the candidates either in the command line. Then the participant would verify the alignments and add decision labels such as insert, ignore, or defer. Finally, the participant would execute another merging script on the verified data to update the in-house graph. To this end, the participant found it convenient to work in the computational notebook setting. The widget-based setup enabled the participant to stitch together three different tasks in the same ecosystem: being able to code, interactively explore, extract data from their interaction, and then execute follow-up programmatic decision --- ``$\ldots$ writing code, recompile, (and) visualize; this interactive and graphical feature is much better than CLI-based setup.'' 

\stitle{Ease of Error Detection.} In particular, the participant appreciated the ability to view the alignment candidates in the context of the corpus and graph (as shown by the bottom two components in Figure~\ref{fig:alignment-verification}) and make decisions more confidently. The participant observed one alignment candidate\footnote{We are refraining from providing the specific example due to constraints related to proprietary data.} where the corpus mention of an entity and the name of the nodes were similar but captured very different semantics when their context was viewed --- ``the structural property of the node and the semantics of the extraction (corpus entity) are not the same.''

\section{Discussion}
\label{sec:discuss}
We acknowledge that the scope of our evaluation is limited, specifically targeting the graph exploration domain in the broader landscape of data science workflows while focusing on an industrial setting. Moreover, all the case study participants were employed at a single company. 
While we believe the observed practices
exist widely in industry and academia, the choice of participants inevitably impacted the generalizability of the findings due to organizational norms, policies, and infrastructures. Additional studies could explore \system's usage benefits and limitations in diverse settings.
Even so, the observed usage benefits and limitations of case studies highlight several opportunities for improvement. 

\stitle{Graph semantics and generalizability.} In this work, we incrementally added new components and interactions based on feedback from participatory sessions. As \system is adopted more and more within \company, continuous addition of graph visualizations, composite interfaces, and data operations would be required. However, these additions should not be arbitrary and practices should be enforced to ground their design on existing graph visualization and graph task taxonomy research. One intriguing aspect of \system framework is its potential for generalizing to other domains beyond graphs, which has already been demonstrated with the table widgets. Supporting a new domain can be achieved by adding data management capabilities for the corresponding domain such as tables, text, and image. Similar to graphs, enabling exploratory analysis in other domains would require modeling the domain-specific semantics. 

\stitle{Addressing the layout gap.}
While not directly a limitation of \system, the linear layout of notebooks may attenuate the benefits afforded by \system. \system attempts to bridge the layout gap by enabling multiple coordinated visualization and resizable panes. Even so, as the analytic session progresses, the number of widgets added may increase. Future studies may explore benefits of approaches such as split panes for coding and visualization (similar to B2~\cite{wu2020b2} and Leam~\cite{rahman2020leam}) or sticky views such as StickyLand~\cite{wang2022stickyland}, in the context of \system widgets.


\section{Conclusion}
\label{sec:conclusion}
We contribute \system, an interactive widget suite
for composing  
that enables knowledge acquisition and integration tasks
within computational notebooks. 
Compared to participants' previous experience,
\system widgets
enabled efficient executions of these tasks
by blending programming and interactive analysis.
Such a design provided a seamless experience to the 
users by eliminating tedious context switching.
Moreover, the multiple coordinated view-based
design of the widgets helped users 
uncover data quality issues
and effectively review agent recommendations. 
In the short term, we continue to employ \system
to meet the needs of data practitioners in \company.
We plan to deploy \system to other workflows within
the HRKG platform within \company which would require
additional need finding and participatory design of widgets
and enhancement of the graph service library.


\bibliographystyle{ACM-Reference-Format}
\bibliography{paper}

\begin{appendix}
\section{\system Components Design via Participatory Sessions}
\label{app_sec:components}
\begin{figure*}[!htb] 
  \centering 
  \includegraphics[width=0.8\linewidth,trim={50 35 40 15},clip]{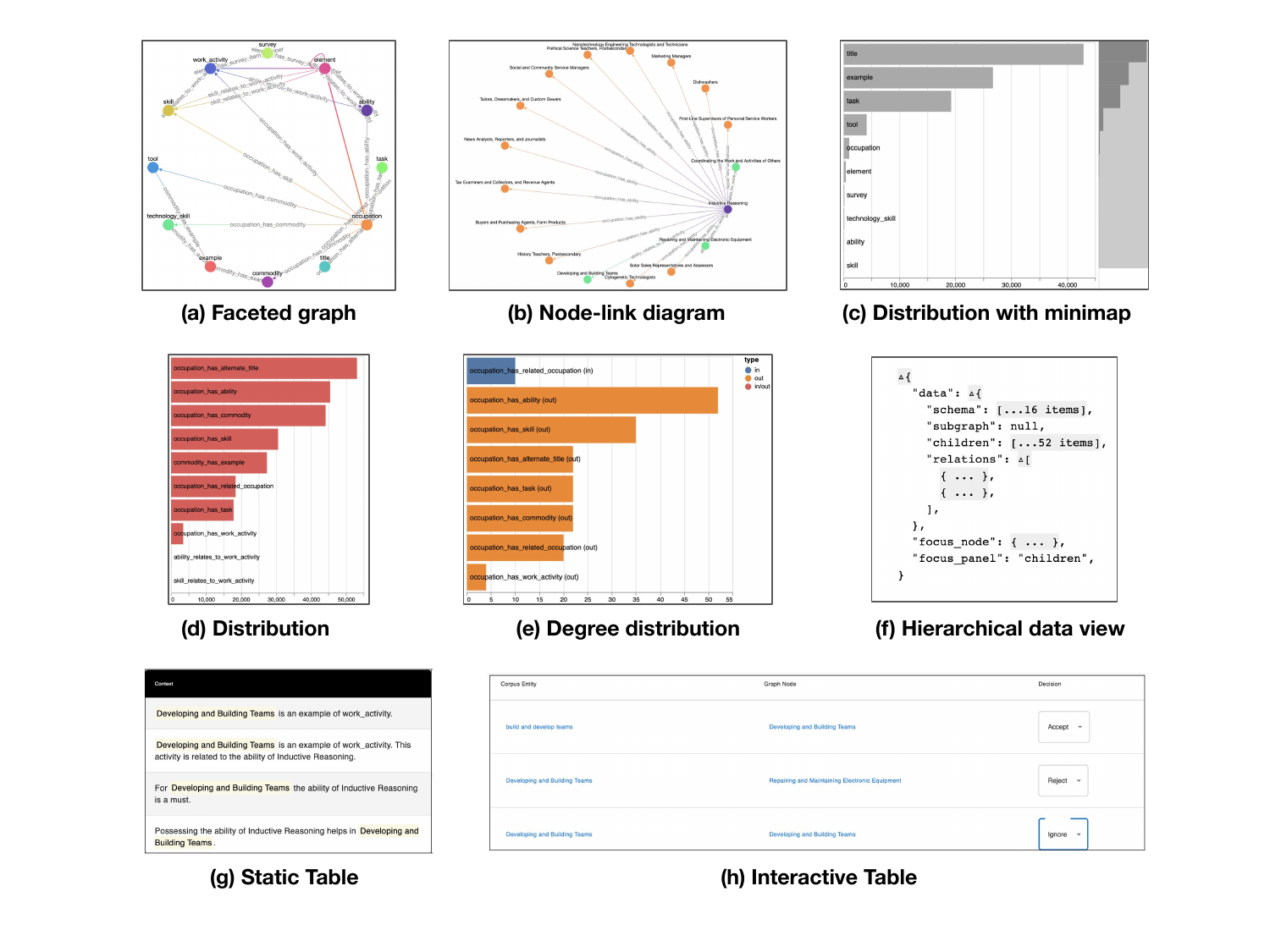}
  \caption{\system base components: (a) faceted graph, (b) node-link diagram, (c) distributions with minimap, (d) distributions, (e) relation degree distributions, (f) hierarchical data viewer, (g) static table, and (h) interactive table.}
  \label{fig:defaults} 
\end{figure*}

In this section, we provide a detailed account of how the participatory session with participants helped prioritize the select of these components (see Figure~\ref{fig:defaults}.)

\stitle{Faceted Graph for Seamless Overview.} A common task encountered across projects was free-form graph exploration --- team members would explore graphs in graphical interfaces, \eg Neo4j browsers, to understand the schema and get a high-level overview of the application domain. They would write scripts to compute entity type and relation distributions and filter the distributions via specific entity type or relation. ll of the projects involved working with large-scale graphs and text corpus. While \system defers computation of all of the data operations in the back-end, there were additional scalability challenges, both perceptual and interactive~\cite{rahman2021noah,liu2013immens}. For example, node-link diagrams do not perceptually scale to thousands of nodes and edges, resulting in visual clutter. In addition, the browser memory becomes a bottleneck when rendering node-links diagrams for large graphs using a force-directed layout. Participants encountered such issues when exploring graphs in web-based Neo4j graph browsers.
All of these tasks can be characterized as overview operations. We created component (a), known as a faceted graph or super-graph~\cite{2019_eurovis_mvnv}, to generate overview of large graphs.
We created component (d) to display entity type and relation distributions. 
However, scrolling the distributions involving too many elements, \eg nodes, and relations, can be cumbersome. So we implemented component (c), a distribution with a minimap, to get a high-level overview of the distribution and scroll to regions of interest (similar to text editors such as sublime text.) 

\stitle{Multiple Coordinated Views to Reduce Visual Discontinuity.} Besides new base components, participants also requested combining multiple components within a widget to accomplish more advanced tasks. The nature of the free-form exploration necessitated such views to first get an overview and then get additional details on demand. For example, participants requested multiple-coordinated views between graph and text corpus for two projects to obtain additional details on data points of interest. Figrue~\ref{fig:alignment-verification} shows an example of such a widget. Similarly, when clicking a node in the faceted graph or a bar in node distribution, participants requested to classify the relation distributions based on their incidence direction on the node, \ie indegree or outdegree. So we implemented component (e) to display relation distribution by their incidence direction corresponding to a node.

\begin{figure*}[!htb] 
  \centering
  \includegraphics[width=0.8\linewidth]{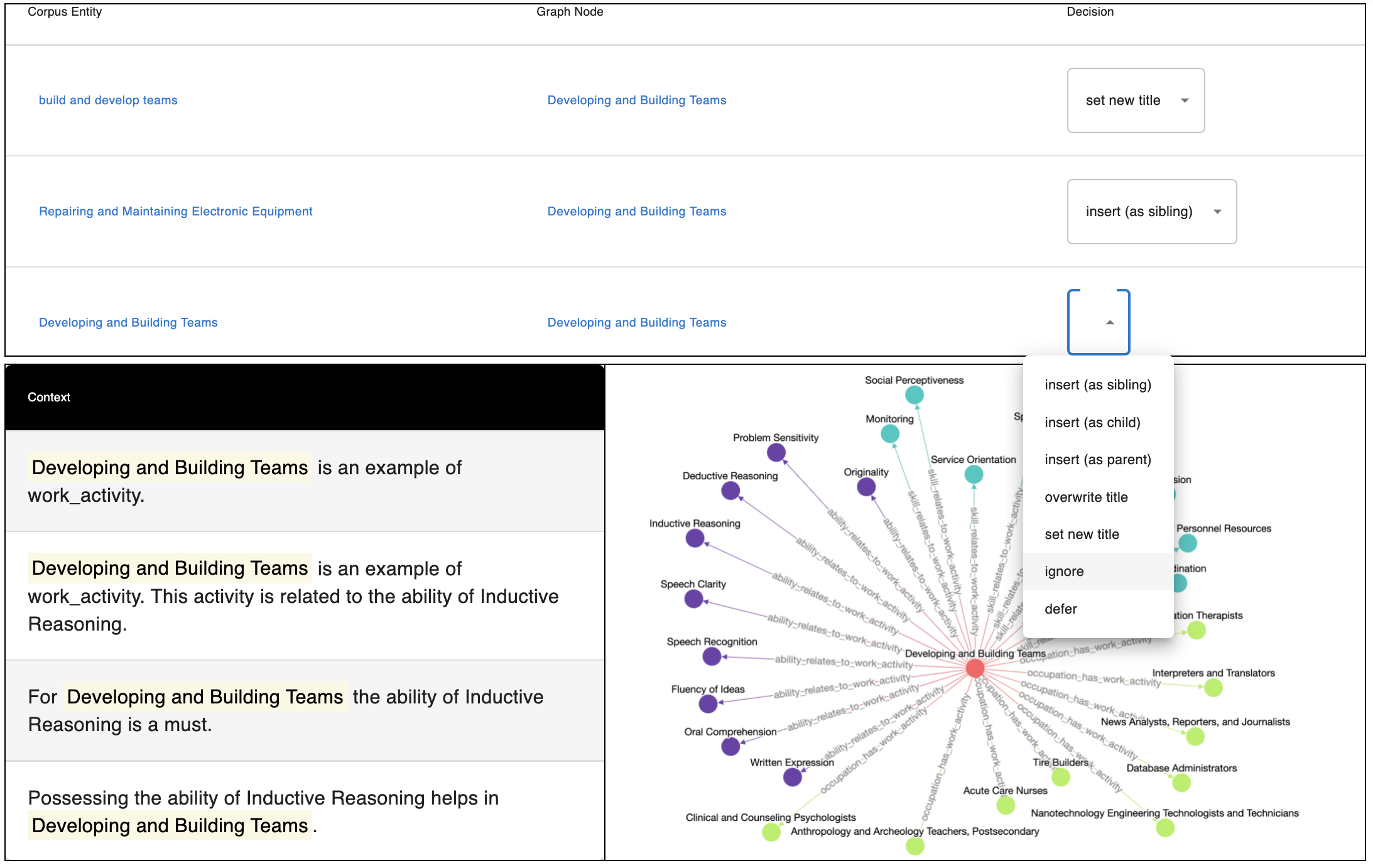}
  \caption{Verification widget for reviewing knowledge integration candidates. Experts may choose a suitable option that reflect their reviewing decision.}
  \label{fig:alignment-verification} 
\end{figure*}

\stitle{Affordances for Multi-modal Data Exploration.} As mentioned earlier, two of the projects involved exploring a text corpus alongside the graph to obtain additional context for decision making. In their current setup, practitioners used spreadsheets for viewing the corpus and manually filter relevant data. So they requested a mechanism to view relevant entries from the text corpus, \eg textual descriptions of entities (nodes) in the graph. In response, we developed component (h), a static table with spans of interest (\eg graph entity) highlighted in the text. Additionally, for one of the projects, participants requested advanced rendering functionalities in table cells (\eg buttons, selection menu.) We developed component (g), an interactive table, to support those features.
We implemented component (f), a hierarchical data viewer, to view JSON objects such as data states and items returned by data accessors. Participants requested this feature since the other alternatives were cumbersome, either loading the data in a JSON formatting application or viewing non-collapsible JSON outputs in a notebook cell.

\end{appendix}

\end{document}